\documentstyle[amssymb,aps,psfig]{revtex}
\begin{document}

%
%  The actual letter
%
%
%
\newcommand {\pp}       {\mbox{p-p}}
\newcommand {\pn}       {\mbox{p-n}}
\newcommand {\pA}       {\mbox{p-A}}
\newcommand {\pBe}      {\mbox{p-Be}}
\newcommand {\pCu}      {\mbox{p-Cu}}
\newcommand {\pAu}      {\mbox{p-Au}}
\newcommand {\SiAu}      {\mbox{Si-Au}}
\newcommand {\AuAu}      {\mbox{Au-Au}}
\newcommand {\PbPb}      {\mbox{Pb-Pb}}
\newcommand {\AGeV}     {\mbox{A$ \cdot$GeV/c}}
\newcommand {\avgdeltay} {\mbox{$\langle \Delta y \rangle$}}
\newcommand {\AvgnuA}   {\mbox{$\langle \nu \rangle_A$}}
\newcommand {\Avgnungrey}   {\mbox{$\bar{\nu} (N_{\rm grey})$}}
\newcommand {\Cher}     {Cherenkov}
\newcommand {\dedx}     {\mbox{$dE/dx$}}
\newcommand {\avgdedx}  {\mbox{$\langle dE/dx \rangle$}}
\newcommand {\sumdedx}  {\mbox{$\sum \langle dE/dx \rangle$}}
\newcommand {\deltay}   {\mbox{$\Delta y$}}
\newcommand {\etal}     {{\it et al.}}
\newcommand {\gammaconv}{\mbox{$\gamma \rightarrow e^{+}e^{-}$}}
\newcommand {\epem}     {\mbox{$e^+e^-$}}
\newcommand {\Et}       {\mbox{$E_t$}}
\newcommand {\fourpi}   {\mbox{$4\pi$}}
\newcommand {\Kp}       {\mbox{K$^+$}}
\newcommand {\Ktopi}    {\mbox{K$/\pi$}}
\newcommand {\Km}       {\mbox{K$^-$}}
\newcommand {\Ks}       {\mbox{K$_S$}}
\newcommand {\Kz}       {\mbox{K$^0$}}
\newcommand {\Kbar}     {\mbox{$\rm \bar{K}$}}
\newcommand {\Kzbar}    {\mbox{$\rm \bar{K^0}$}}
\newcommand {\kkbar}    {\mbox{${\rm K\bar{K}}$}}
\newcommand {\Ksdecay}  {\mbox{K$^0_S \rightarrow \pi^+\pi^-$}}
\newcommand {\Lam}      {\mbox{${\rm \Lambda}$}}
\newcommand {\Lamdecay} {\mbox{${\rm \Lambda} \rightarrow p\pi^{-}$}}
\newcommand {\Minv}    {\mbox{$M_{\rm inv}$}}
\newcommand {\mperp}    {\mbox{$m_{\perp}$}}
\newcommand {\Ngrey}    {\mbox{$N_{\rm grey}$}}
\newcommand {\NN}       {\mbox{N-N}}
\newcommand {\Npart}    {\mbox{$N_{\rm part}$}}
\newcommand {\nubar}    {\mbox{$\bar{\nu}$}}
\newcommand {\nungrey}  {\mbox{$\nu(N_{\rm grey})$}}
\newcommand {\Nnu}      {\mbox{$\nu$}}
\newcommand {\Npim}     {\mbox{$N_{\pi^-}$}}
\newcommand {\pbar}     {\mbox{$\bar{p}$}}
\newcommand {\pim}      {\mbox{$\pi^-$}}
\newcommand {\pion}     {\mbox{$\pi$}}
\newcommand {\pip}      {\mbox{$\pi^+$}}
\newcommand {\pippim}   {\mbox{$\pi^+\pi^-$}}
\newcommand {\pperp}    {\mbox{$p_\perp$}}
\newcommand {\ppim}     {\mbox{$p\pi^-$}}
\newcommand {\pz}    	{\mbox{$p_z$}}
\newcommand {\sig}      {\mbox{$\sigma$}}
\newcommand {\Sigz}      {\mbox{$\Sigma^0$}}
\newcommand {\sumpt}    {\mbox{$\sum p_\perp$}}
\newcommand {\sumpz}    {\mbox{$\sum p_z$}}
\newcommand {\rap}      {\mbox{$y$}}
\newcommand {\rhoN}     {\mbox{$\rho_{\rm N}$}}
\newcommand {\sigNN}    {\mbox{$\sigma_{\rm NN}$}}
\newcommand {\sqrts}    {\mbox{$\sqrt{s}$}}
\newcommand {\usec}     {\mbox{$\mu s$}}
\newcommand {\ylead}    {\mbox{$y_{\rm lead}$}}
\newcommand {\ynn}   	{\mbox{$y_{\rm NN}$}}
\newcommand {\dndy}     {\mbox{$dn/dy$}}
\newcommand {\mt}       {\mbox{$m_{\perp}$}}
\newcommand {\mtmo}     {\mbox{$m_{\perp}-m_0$}}
\newcommand {\VO}       {\mbox{$V_0$}}
\newcommand {\chisq}    {\mbox{$\chi^{2}$}}
\newcommand {\normchisq}{\mbox{$\chi^{2}_{V_0}/NDF$}}

\title{
Semi-Inclusive \Lam\ and \Ks\ Production in p-Au Collisions at 17.5~GeV/c
}

%
%	authors
%

\author{
	I.~Chemakin$^{2}$,
	V.~Cianciolo$^{7,8}$,
	B.A.~Cole$^{2}$,
	R.~Fernow$^{1}$,
	A.D.~Frawley$^{3}$,
	M.~Gilkes$^{9}$,
	S.~Gushue$^{1}$,
	E.P.~Hartouni$^{7}$,
	H.~Hiejima$^{2}$,
	M.~Justice$^{5}$,
	J.H.~Kang$^{11}$,
	H.~Kirk$^{1}$,
	N.~Maeda$^{3}$,
	R.L.~McGrath$^{9}$,
	S.~Mioduszewski$^{10}$,
	D.~Morrison$^{10,1}$,
	M.~Moulson$^{2}$,
	M.N.~Namboodiri$^{7}$,
	G.~Rai$^{6}$,
	K.~Read$^{10}$,
	L.~Remsberg$^{1}$,
	M.~Rosati$^{1,4}$,
	Y.~Shin$^{11}$,
	R.A.~Soltz$^{7}$,
	S.~Sorensen$^{10}$,
	J.H.~Thomas$^{7,6}$,
	Y.~Torun$^{9,1}$,
	D.~Winter$^{2}$,
	X.~Yang$^{2}$,
	W.A.~Zajc$^{2}$,
	and Y.~Zhang$^{2}$,
	}

%
%	institutions
%
\bigskip
\address{
$^{1}$ Brookhaven National Laboratory, Upton, New York 11973\\
$^{2}$ Columbia University, New York, NY 10027, 
$^{3}$ Florida State University, Tallahassee, FL 32306, \\
$^{4}$ Iowa State University, Ames, IA 50010, $^{5}$ Kent State University, Kent, OH 44242\\
$^{6}$ Nuclear Science Division, Lawrence Berkeley National Laboratory, Berkeley, CA 94720 \\
$^{7}$ Lawrence Livermore National Laboratory, Livermore, CA 94550 \\
$^{8}$ Oak Ridge National Laboratory, Oak Ridge, TN 37831\\
$^{9}$ State University of New York at Stony Brook, Stony Brook, NY 11794\\
$^{10}$ University of Tennessee, Knoxville, TN 37996, $^{11}$ Yonsei University, Seoul 120-749, Korea \\
}

\date{\today}
\maketitle
\bigskip
\begin{abstract}
%
% Note: 600 characters max !!!
%
%
The first detailed measurements of the centrality dependence of 
strangeness production in \pA\ collisions are presented. 
\Lam\ and \Ks\ $dn/dy$ distributions from 17.5~GeV/c p-Au 
collisions are shown as a function of ``grey'' track multiplicity 
and the estimated number of collisions, \Nnu, made by the 
proton. The \Nnu\ dependence of the \Lam\ yield deviates from a 
scaling of \pp\ data by the number of participants, increasing faster
than this scaling for $\nu \leq 5$ and saturating for larger \Nnu.
A slower growth in \Ks\ multiplicity with \Nnu\ is
observed, consistent with a weaker \Nnu\ dependence of \kkbar\
production than $YK$ production. 

\end{abstract}

\twocolumn

%% @(#) $Id: body.tex,v 1.8 2000/06/08 20:13:56 cole Exp $
%% 
%% Created by: Brian A. Cole
%% Creation date: Long time ago
%% 
%% Description:
%%     Lambda and Kshort PRL paper
%% 
%% Log Message: $Log: body.tex,v $
%% Log Message: Revision 1.8  2000/06/08 20:13:56  cole
%% Log Message: Updates to NEW repository for changes in response to referee's 1st comments
%% Log Message:
%% Log Message: Revision 1.7  2000/02/29 10:08:46  cole
%% Log Message: Nearly last edit of paper
%% Log Message:
%% Log Message: Revision 1.6  2000/01/20 19:30:03  cole
%% Log Message: Finished full draft of paper
%% Log Message:
%% Log Message: Revision 1.5  1999/12/14 05:47:33  xhyang
%% Log Message: *** empty log message ***
%% Log Message:

Significant effort has been devoted in the last decade to 
measuring strange particle production in nucleus-nucleus (A-A) collisions,
motivated by predictions that quark-gluon plasma (QGP) formation could
enhance strangeness
\cite{Raf82:Strangeness,Koc86:Probing}. Experiments at both the 
Brookhaven National Laboratory (BNL) AGS and the CERN SPS accelerators
have reported large increases in relative strange particle yields in
central light (Si, S)
\cite{Abb90:Kaon,Ahle:1999va,Alber:1994tz,Antinori:1999dy}
and heavy ion (Au, Pb)
\cite{Ahle:1998gv,Ahm96:Lambda,Bormann:1997qa,Margetis:1999ge,Andersen:1999ym}
induced collisions compared to p-p collisions. However, we still
cannot claim a true understanding of the physics of strangeness
enhancement due to the complexity of the hadronic interactions
underlying A-A collisions and the competing mechanisms
proposed and/or used in models
\cite{Mat89:K,Pan92:Cascade,Braun-Munzinger:1996mq} to explain the data. 
%%A recent study of the systematics of kaon production in Si+A
%%and Au+Au collisions at the AGS \cite{Ahle:1999va} suggests that the
%%strangeness yield 
%is significantly enhanced in light-ion collisions
%%% and that this
%%%enhancement saturates in central Si+Au collisions. 

The difficulty in interpreting the A-A data and the observation that
the enhancement is already present in light-ion collisions 
suggest the use of \pA\ collisions to study this problem further.
The simpler final state of \pA\ collisions may allow the production
rate for strange particles to be directly connected with the
scattering dynamics of the incoming proton. Previously
published \pA\ data 
\cite{Abb91:Comparison,Abb92:Measurement} demonstrated an 
increase in the inclusive \Ktopi\ ratio with increasing A at the AGS,
suggesting that the strangeness enhancement mechanism is already at
work in \pA{} collisions. A more thorough analysis
\cite{Gazdzicki:1994rs} of inclusive data at higher energies
\cite{Bialkowska:1992kq} suggested no overall strangeness enhancement
but a possible modest enhancement in \Lam{} production offset by a
decrease in \Kbar{} production. This result has led to claims that the
observed strangeness enhancement in A-A collisions at SPS energies may 
result from QGP formation \cite{Gazdzicki:1997hm}. However,
extrapolations to central A-A data from 
{\em inclusive} p-A data are intrinsically flawed: Centrally selected
A-A events necessarily involve more scatterings of the participant
nucleons, and the dynamics of strangeness production may be quite
sensitive to these additional scatters. A resolution of this problem 
requires a detailed measurement of the centrality dependence of strangeness
production in \pA\ collisions.
%A first attempt at such an analysis
%was reported in \cite{Bri92:PA}, but the low statistics of this
%analysis prevent significant conclusions from being drawn.

In this paper we present the first such measurement, made by BNL
experiment 910 at the AGS accelerator. The data consist of \Lam{} and
\Ks{} rapidity spectra and integrated yields obtained as a function of
``grey'' track multiplicity from \pAu\ collisions at a beam momentum
of 17.5~GeV/c. E910's nearly complete rapidity coverage allows us to
accurately estimate the total \Lam{} and \Ks{} multiplicities and
study the variation of absolute yields with $\nu(N_{\rm grey})$, the
estimated number of collisions of the proton in the Au nucleus. 
A common benchmark \cite{Ahle:1999va,Andersen:1998vu,Andersen:1999ym}
for evaluating strangeness enhancement in A-A collisions is the scaling of
\pp\ data by the number of participants, \Npart, in the collision. For
p-A collisions, an \Npart\ scaling of \pp\ data would yield~\cite{npart}
\begin{equation}
N_{\rm prod} = \frac{1}{2} N^{\rm pp}_{\rm prod} (1 + \nu),
\label{eq:wn}
\end{equation}
since there are two participants in \pp\ collisions. We compare our
multiplicities to the yields expected from Eq.~1 to evaluate whether
we see enhanced strange particle production using the same benchmark
as in A-A collisions. We observe that for \Lam\ production Eq.~1 is 
physically sensible since baryon number is conserved and ${\rm
B\bar{B}}$ processes are negligible at our energies. Deviations from
Eq.1 would imply contributions from target nucleons not directly
struck by the projectile or changes in the probability of participants
to fragment into \Lam's.
%%%%%%%%%%%%%%%%%%%%%%%%%%%%%%%%%%%%%%%%%%%%%%%%%%%%%%%%%%%%%%%%%%%%%
%We instead compare the %observed yields to expectations from the
%so-called %``wounded-nucleon'' (WN) prescription for particle
%multiplicities, %\begin{equation} %N_{\rm prod} \approx \frac{1}{2}
%N^{\rm pp}_{\rm prod} (1 + \nu), %\label{eq:wn} %\end{equation} %This
%relationship embodies one of the most %critical results from previous
%studies of \pA{} collisions -- the %slow growth with A of particle
%yields at and above the nucleon-nucleon %center of mass -- and
%accurately describes the total yield of produced %pions in \pA{}
%collisions \cite{Eli80,Dem84:PA}. It is also a useful %benchmark for
%comparison with A-A data where yields are often %evaluated as a
%function of the number of ``participant'' nucleons %(\Npart) since in
%a \pA{} collision $N_{\rm part} = 1+\nu$.

E910 was staged in the Multi-Particle Spectrometer (MPS) facility at
the AGS. For this data, the Cherenkov tagged secondary proton beam had a mean momentum of $17.5 \pm 0.1$~GeV/c and a 
1.5\% momentum spread. The E910 spectrometer was previously described in 
\cite{Chemakin:1999cp,Che99:Slow-Proton}; the results presented here 
rely on the EOS time projection chamber and the beamline/trigger counters.
%The TPC was located at the approximate center of the pole-tip of
%the MPS ``C'' magnet with the 120~V/cm drift field aligned with the nominal
%0.5~T primary (vertical) component of the magnetic field. 
%
%It was filled with P10 gas at atmospheric pressure and read out on 15360 pads
%arranged in 128 rows down the 1.5~m long-axis of the TPC. 
%
A 3.9~${\rm g/cm^2}$ Au target was located 20.5~cm upstream of the
active area of the TPC and was immediately followed by a ${\rm 10~cm
\times 10~cm}$ two-layer scintillating-fiber hodoscope. This
multiplicity detector provided two triggers, a minimum-bias (MB)
trigger that required 2 hits on 
each layer and a ``central'' trigger that required a total of 20 hits
in the two layers and selected approximately the 25\% most central
events. Due to the low light yield of the fibers the MB trigger 
suffered significant efficiency loss for low multiplicity events with no
highly ionizing tracks.

\begin{figure}[!t]
\centerline{
\hbox to \hsize{
%\begin{tabular}{ccc}
%\psfig{file=acceptance.eps,height=1.2in,bbllx=25pt,bblly=0pt,bburx=490pt,bbury=350pt}
%\hfill
%\psfig{file=lambda_invmass.eps,height=1.3in,bbllx=2pt,bblly=15pt,bburx=195pt,bbury=350pt} 
%\hfill
%\psfig{file=kshort_invmass.eps,height=1.3in,bbllx=2pt,bblly=15pt,bburx=195pt,bbury=350pt} 
\psfig{file=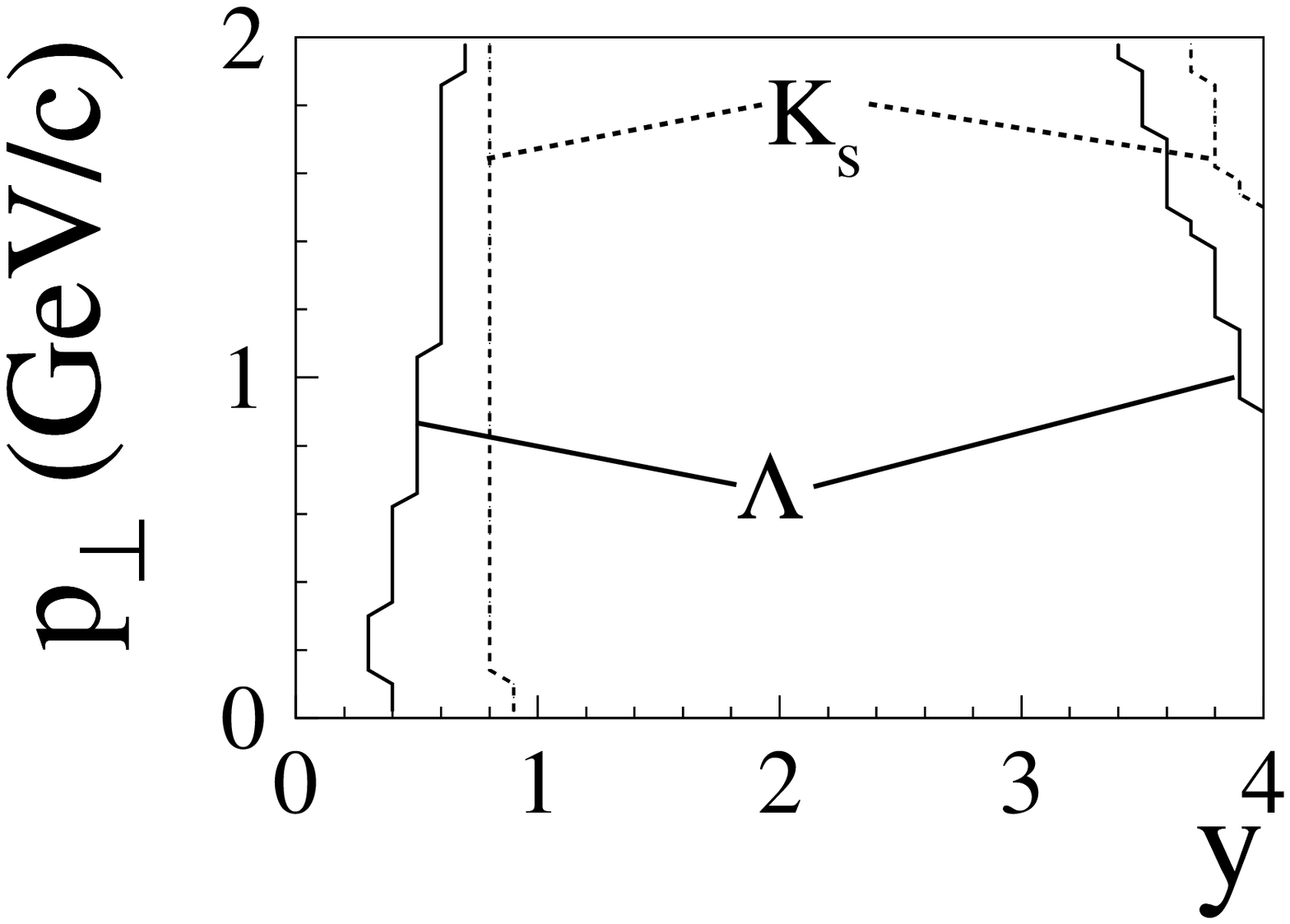,height=1.3in,bbllx=25pt,bblly=0pt,bburx=490pt,bbury=350pt}
\hfill
\psfig{file=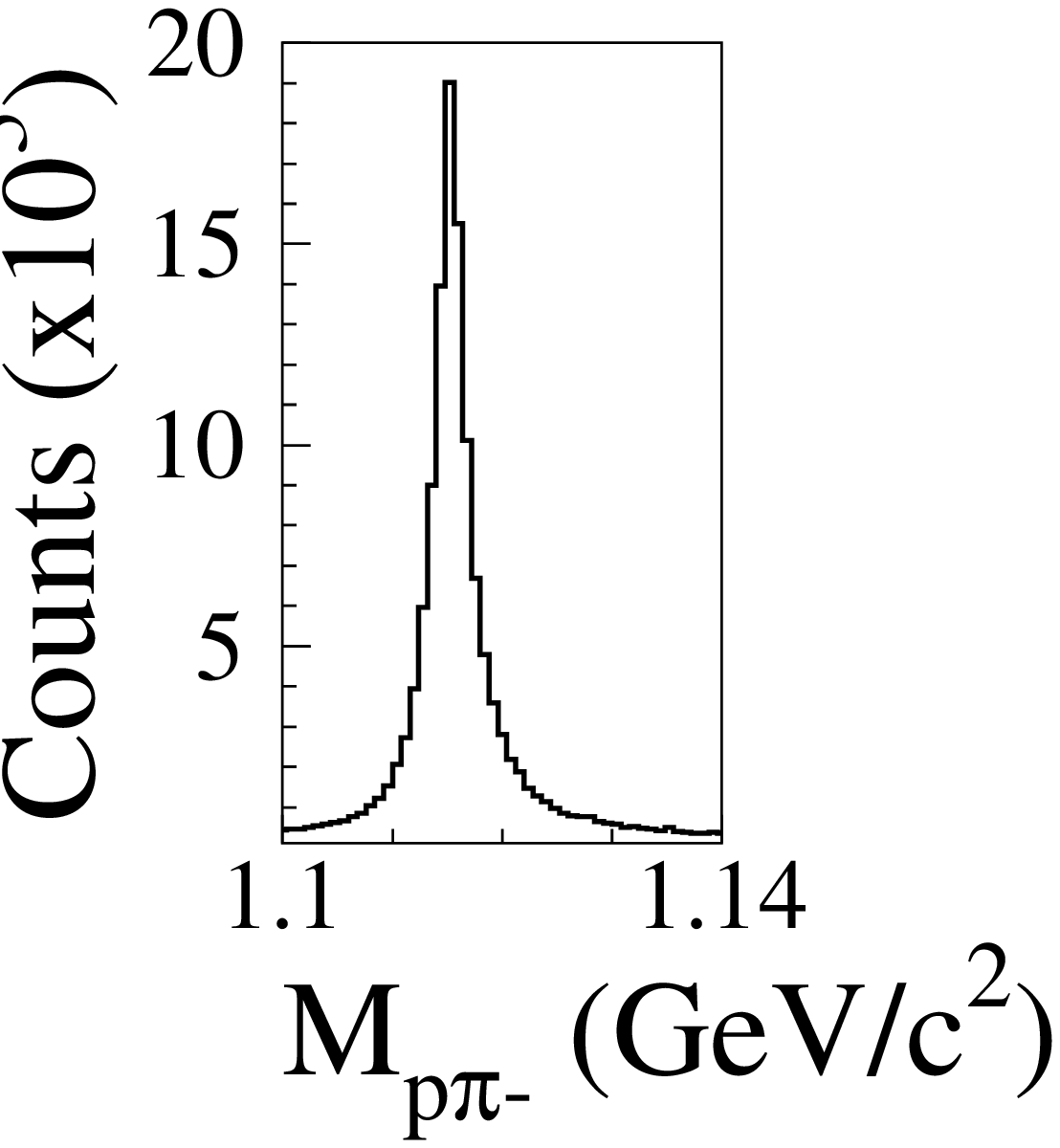,height=1.3in,bbllx=10pt,bblly=15pt,bburx=185pt,bbury=350pt} 
\hfill
\psfig{file=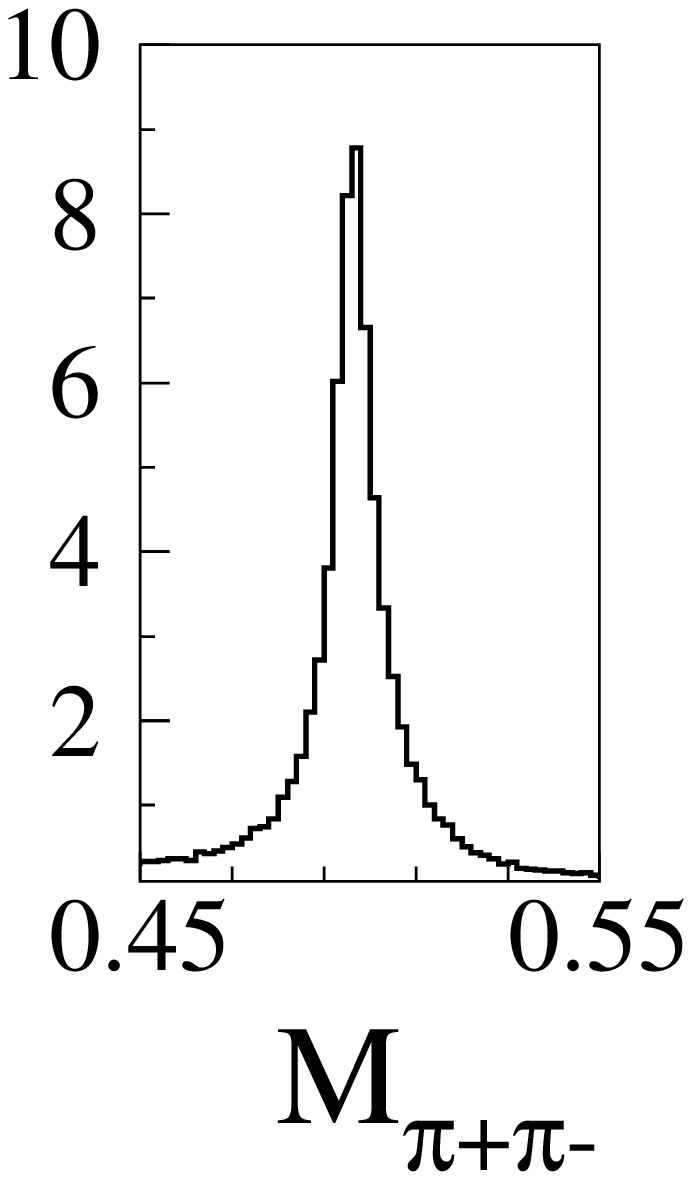,height=1.3in,bbllx=10pt,bblly=15pt,bburx=185pt,bbury=350pt} 
%\end{tabular}
}}

\caption{\Lam{} and \Ks{} acceptance boundary (left) and
reconstructed invariant mass spectra (right).}
\label{fig:lamKsAccept}
\end{figure}

After finding and fitting the recorded pulses in the TPC we obtained
typical resolutions of 0.7~mm (vertical) and 0.5~mm (horizontal) for
position measurements in each sample. The momentum resolution for
particles bending in the 0.5~T magnetic field varied from 1.2\% ($p <$
2~GeV/c) to 5.5\% ($p \sim$ 17~GeV/c). The \dedx{} measurement, obtained from
a truncated-mean of the TPC samples on each track, provided a 
resolution of \sig{}$/$\avgdedx{} $= 6\%$ for typical track lengths.
\Lam{}'s and \Ks{}'s were measured and identified through a combination of
topological reconstruction and \dedx\ identification of the decay
daughters. We paired and removed conversion electrons and positrons,
the dominant source of background, with an efficiency of $\approx 50\%$. 
%The dominant source of  backgrounds under the \Lam{} and \Ks{} peaks result
%primarily from mis-identified electrons and positrons; $\approx{}
%50\%$ of this background was removed through reconstruction and
%removal of $e^+e^-$ conversion pairs
%%%%\cite{Wie92:Coulomb-effect}. 
We further reduced background from conversions and from false vertices by
applying tighter geometric cuts to small opening-angle pairs.
We attempted topological fits on $+$/$-$ track pairs satisfying
applied geometric and \dedx\ cuts and accepted as $\rm V_0$ candidates those
passing applied \chisq\ cuts with origin $\geq 3.5~cm$ from the target.
%by requiring that
%small opening-angle pairs have at least one of the daughter tracks not
%associated with the primary vertex and by rejecting decays with one of
%the daughter tracks had lab polar angles $\theta > 70^\circ$. 
%
We used a combined likelihood from the single track \dedx's and 
hypothetical \ppim{} and \pippim{} invariant masses (\Minv) of the
pairs to identify the decaying particle, obtaining the \Minv{}
distributions shown in Fig.~\ref{fig:lamKsAccept}.  We obtained mass
resolutions of FWHM=4~MeV, 13~MeV for \Lam{}'s and \Ks{}'s,
respectively and corresponding S/B ratios of 35:1 and 30:1. The
acceptances were calculated via GEANT simulations of the detector
response to 20M pure \Lam\ and 10M pure \Ks\ decays
\cite{accept_note}. We show in Fig.~\ref{fig:lamKsAccept} the
$y-p_\perp$ regions with $> 10\%$ acceptance. For the centrality
measurement, we identified as ``grey'' tracks protons and deuterons in
the momentum ranges [0.25,1.2]~GeV/c and [0.5,2.4]~GeV/c,
respectively. We obtain the multiplicity of grey tracks, \Ngrey,
within our geometric acceptance event-by-event and from this quantity
estimate the mean number of collisions suffered by the beam proton,
\Avgnungrey, using an established technique \cite{Che99:Slow-Proton}.  

The data presented in this paper resulted from a combined 4.65M 
triggers. We required a valid event to have at least one
secondary charged particle in the final state and a
$\Sigma p_\perp > 85$~MeV/c. In addition, we vetoed one and two-track events
containing a high-momentum positive track consistent with a
quasi-elastically scattered proton. After applying quality and the
above interaction cuts we obtained 2.97M events, 1.88M
minimum-bias and 2.07M central. From these,  
we reconstructed a total of 156.8k \Lam's and 76.8k \Ks's. Using
beam-triggered events we determined trigger efficiency corrections 
%
%\markcut{generated uniformly in rapidity and with an exponential fall-off in \mtmo{}
%with an inverse slope of 150~MeV, in GEANT-based Monte Carlo
%simulation with the same cuts used in data analysis. }
%
as a two-dimensional function of charged-particle multiplicity and
\Ngrey{}. The correction for multiplicity 1,2 events with $\Ngrey=0$ is large
($5.1$) while the average correction for all interactions is $1.1$.

\begin{figure}[!h]
\centerline{
\begin{tabular}{c@{\quad}c}
\psfig{file=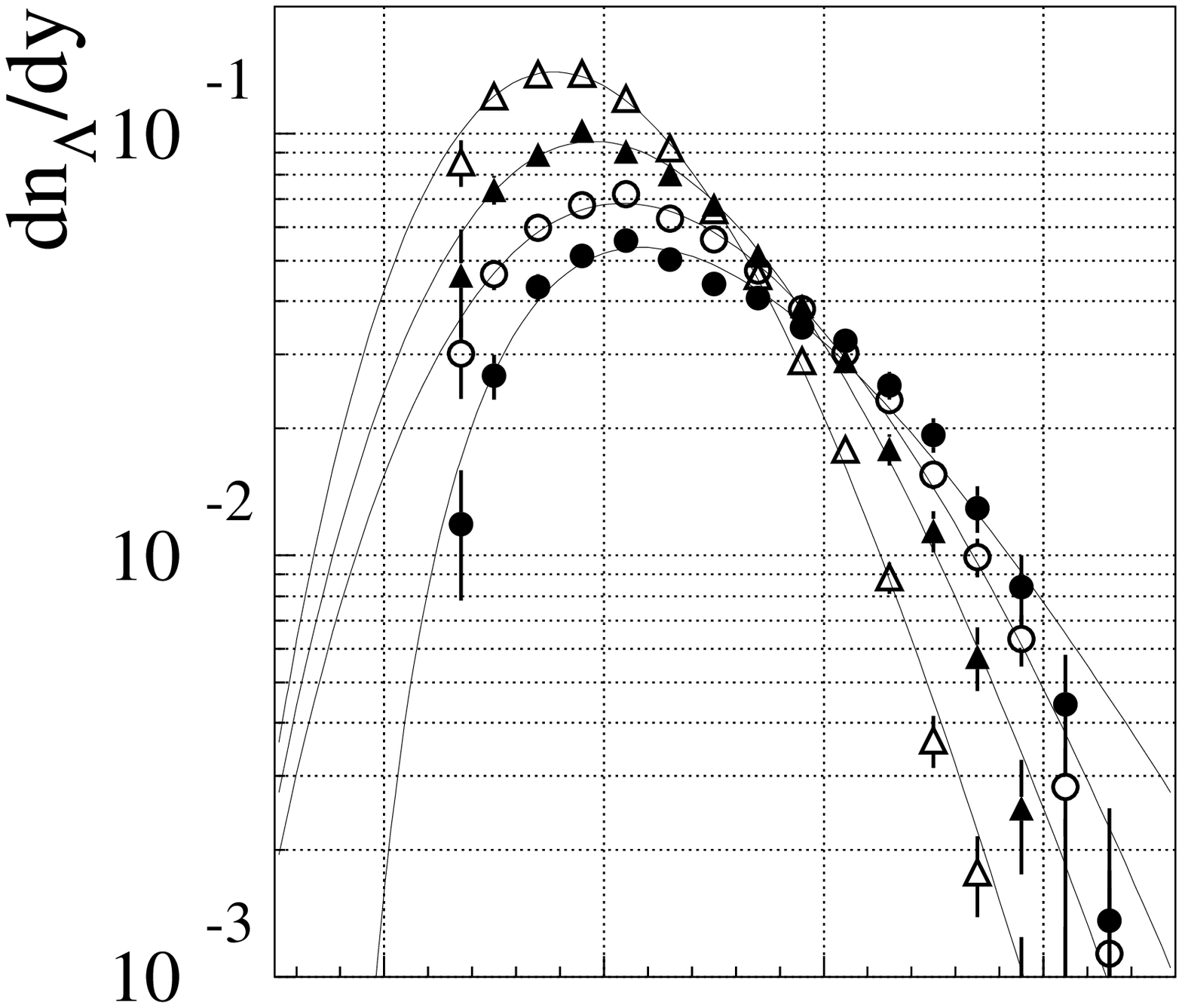,width=1.7in,bbllx=9pt,bblly=83pt,bburx=489pt,bbury=489pt} &
\psfig{file=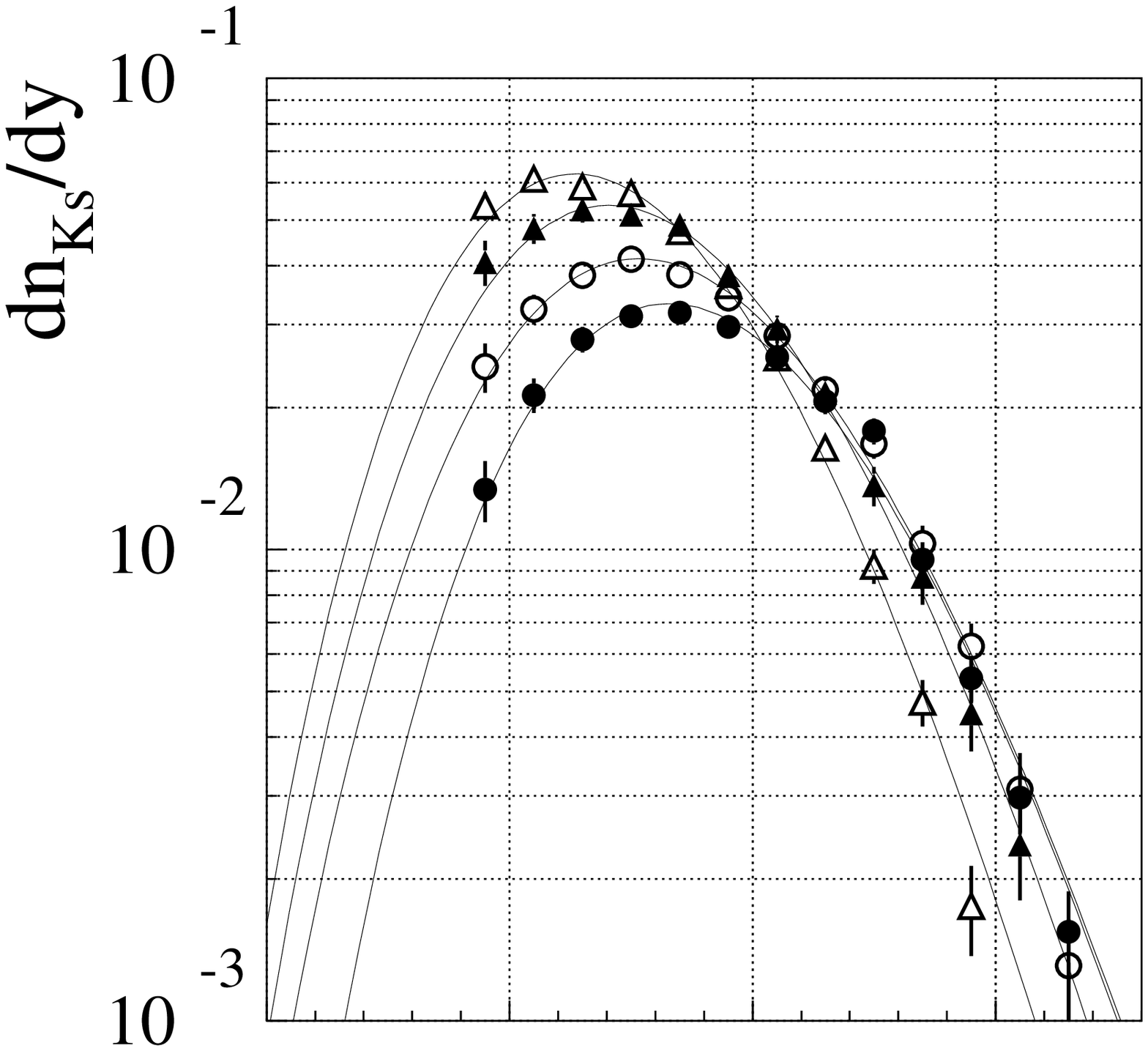,width=1.7in,bbllx=9pt,bblly=83pt,bburx=489pt,bbury=489pt} \\
\psfig{file=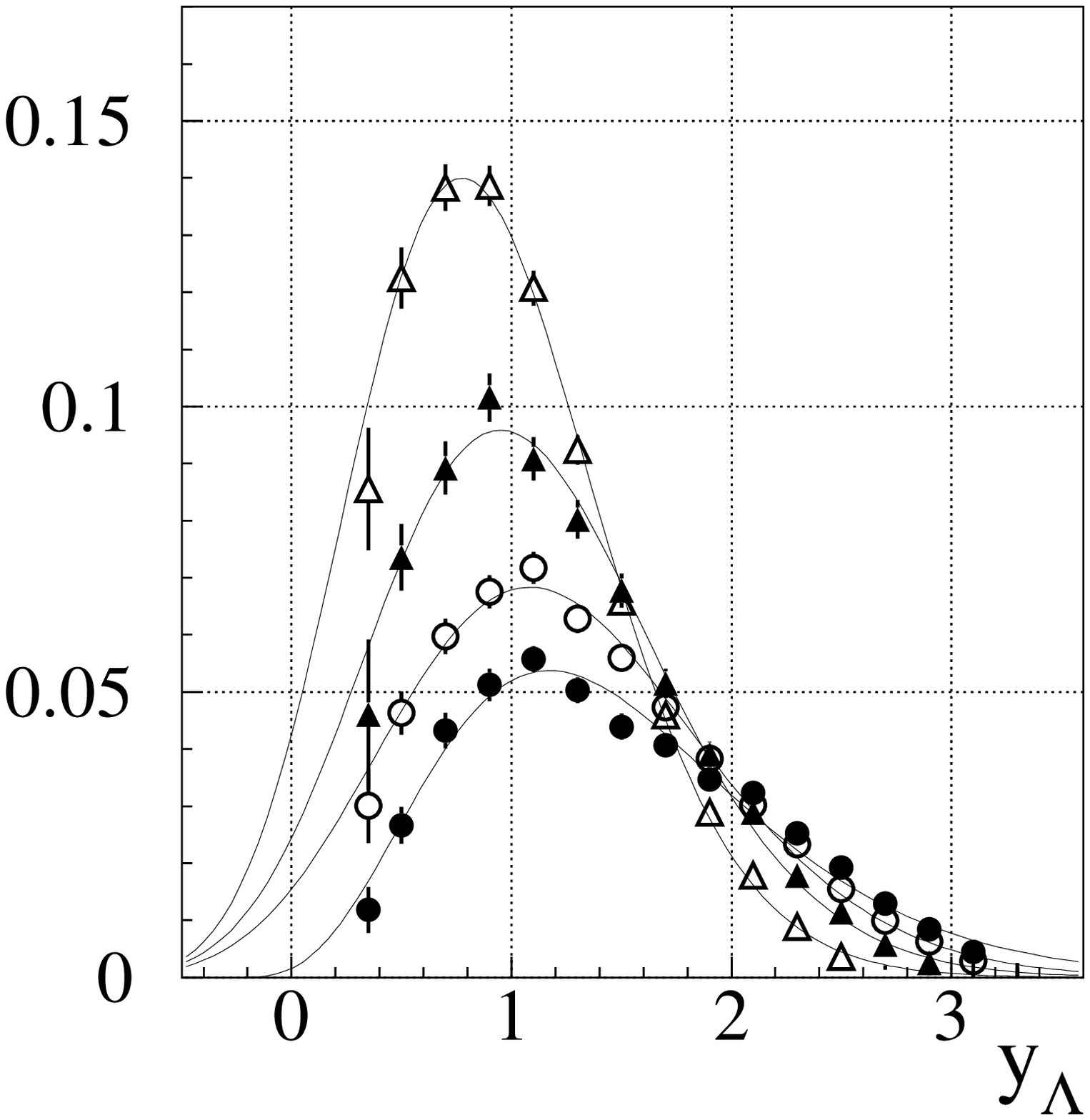,width=1.7in,bbllx=9pt,bblly=22pt,bburx=489pt,bbury=489pt} &
\psfig{file=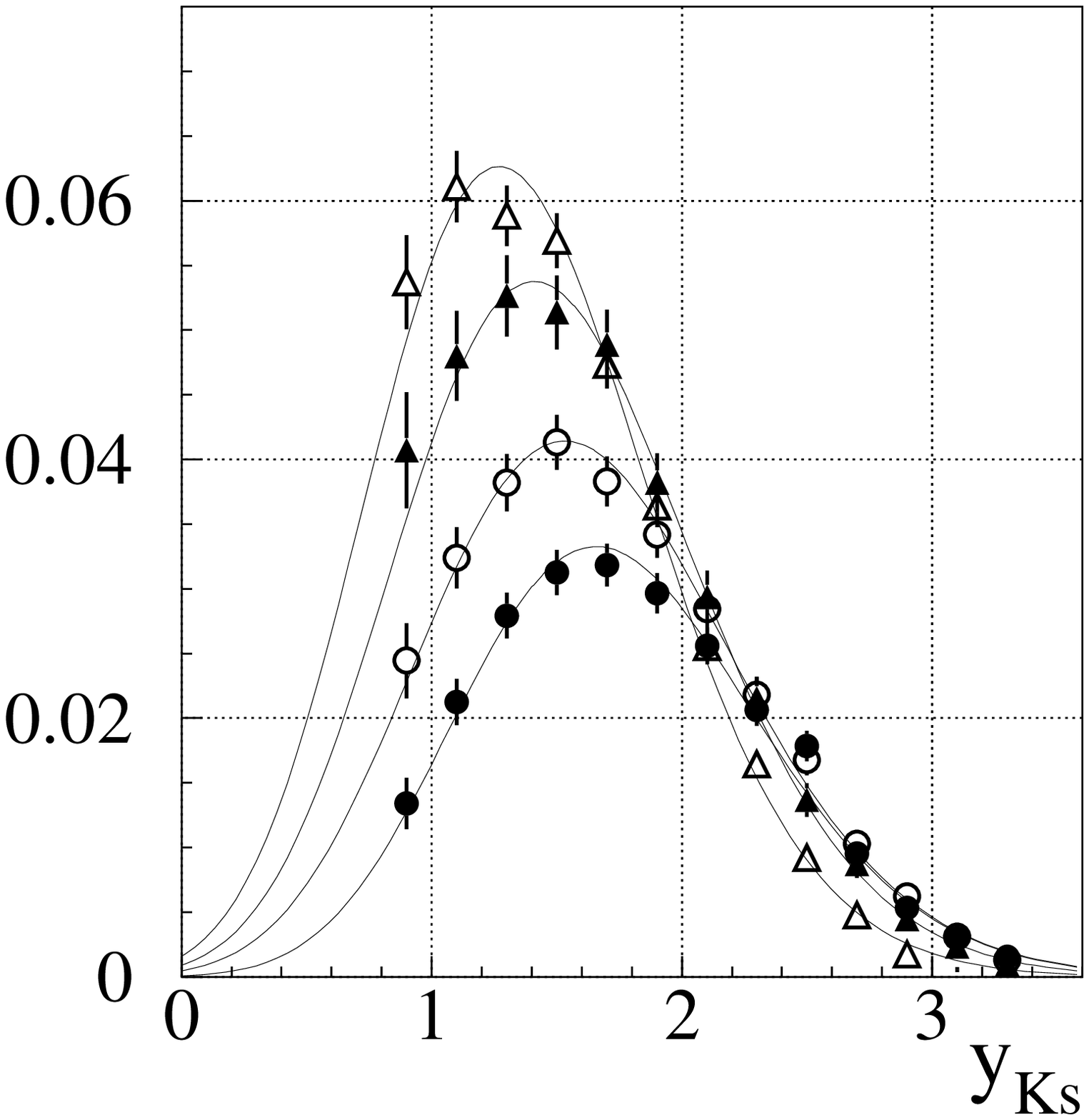,width=1.7in,bbllx=9pt,bblly=22pt,bburx=489pt,bbury=489pt} 
\end{tabular}
}
\caption{\Lam{} (left) and \Ks{} (right) rapidity density
distributions for different \Ngrey{} plotted in both log (top)
and linear(bottom) scale. $N_{\rm grey} = $ $\bullet$ - 0,
$\circ$ - 1, $\blacktriangledown$ - 2 and $\triangledown$ - 4.} 
\label{fig:dndy}
\end{figure}

We calculated \Lam\ and \Ks\ yields $\Delta N(m_\perp, y,\Minv)$ per
event as a function of \Ngrey\ after subtracting the \Lam\ and \Ks\ 
background in each bin. We corrected these for acceptance (A), trigger
efficiency ($\varepsilon$), and branching ratio (BR) to obtain an invariant
differential yield, 
\begin{equation}
\frac{d^2n_{\Lambda/K_s}}{m_\perp dm_\perp dy} = \frac{1}{m_\perp} 
\frac{1}{A \, \varepsilon \, {\rm BR}}
\frac{\Delta
N_{\Lambda/K_s}(m_\perp,y)}{N_{\rm evt} \Delta m_\perp \Delta y  }.
\label{eq:diffyield}
\end{equation}
For $\Ngrey < 4$, we used only MB triggers, but for larger \Ngrey\
we combined both triggers, weighting by the number of events in
each sample. 
%
%We also verified that for these \Ngrey\ bins, the spectra
%obtained separately from the two triggers agree within statistical
%errors for all $(y, m_\perp)$ bins.
%
The \mperp\ spectra are uniformly well-described by exponential
distributions except at the highest \mperp\ values ($\mperp - m > 0.6$
at mid-rapidity) where we do not have sufficient statistics to
accurately determine the background subtraction. We fit the \mperp\
spectra excluding these points to the form, 
\begin{equation}
\frac{1}{2\pi{}m_\perp}\frac{d^2n}{dm_\perp dy} =
\frac{1}{2\pi{}(m_0+T)T} \frac{dn}{dy} e^{-(m_\perp-m_0)/T}.
\label{eq:mtspecfit}
\end{equation}
where \dndy\ is a direct parameter of the fit representing the integral
of Eq.~\ref{eq:mtspecfit} over \mperp. The fits give inverse slopes, $T$,
that vary from 0.05~GeV/c at 
low and high $y$ to 0.14~GeV/c at mid-rapidity, consistent with proton
spectra obtained from \pA\ collisions at a similar energy 
\cite{Abb92:Measurement}. For \Ngrey\
bins where we do not have enough statistics to perform the fits we
directly sum over \mperp\ to obtain \dndy.
%
%The results
%from the $m_\perp$ fits are consistently 5-10\% lower than the results
%from direct summing for all rapidity bins and for both \Lam\ and \Ks. We
%attribute this difference to un-subtracted background at large
%\mperp\ where we can only estimate backgrounds because of low
%statistics. We therefore have chosen to adopt the (lower) \mperp\ fit
%results, and obtain the \dndy\ distributions shown in
%
Fig. \ref{fig:dndy} shows the resulting \dndy\ distributions for a
sub-set of the available \Ngrey\ bins. We estimate 90\%~CL point-to-point
systematic errors in the \Lam\ and \Ks\ \dndy\ measurements, including
contributions from the fitting, to be $< 5\%$ except for the lowest
rapidity bin (20\%), and estimate the normalization systematic error
to be $\pm 10\%$. 

\begin{figure}[!t]
\centerline{
\begin{tabular}{cc}
\psfig{file=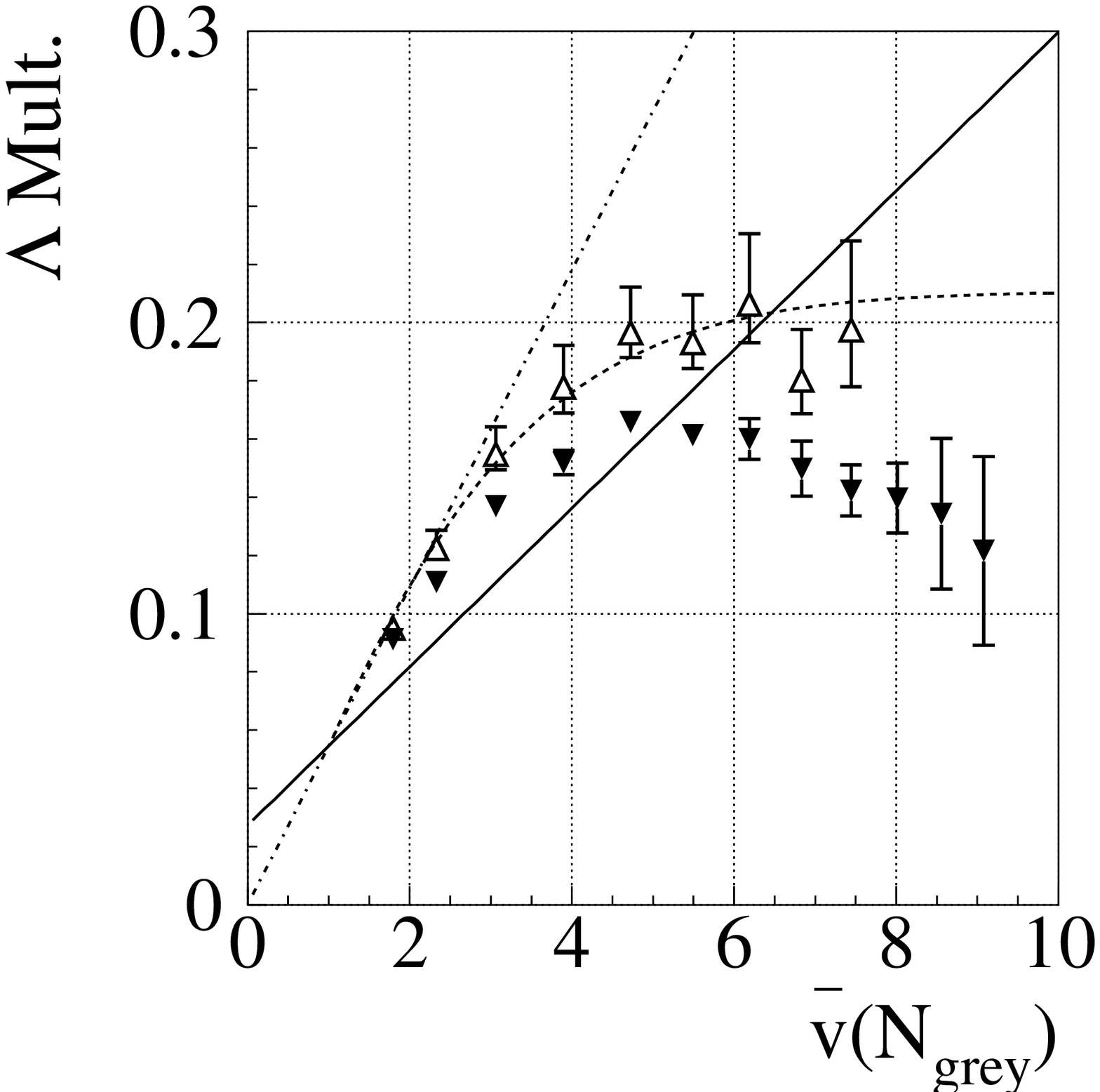,width=1.65in,bbllx=6pt,bblly=5pt,bburx=491pt,bbury=488pt} &
\psfig{file=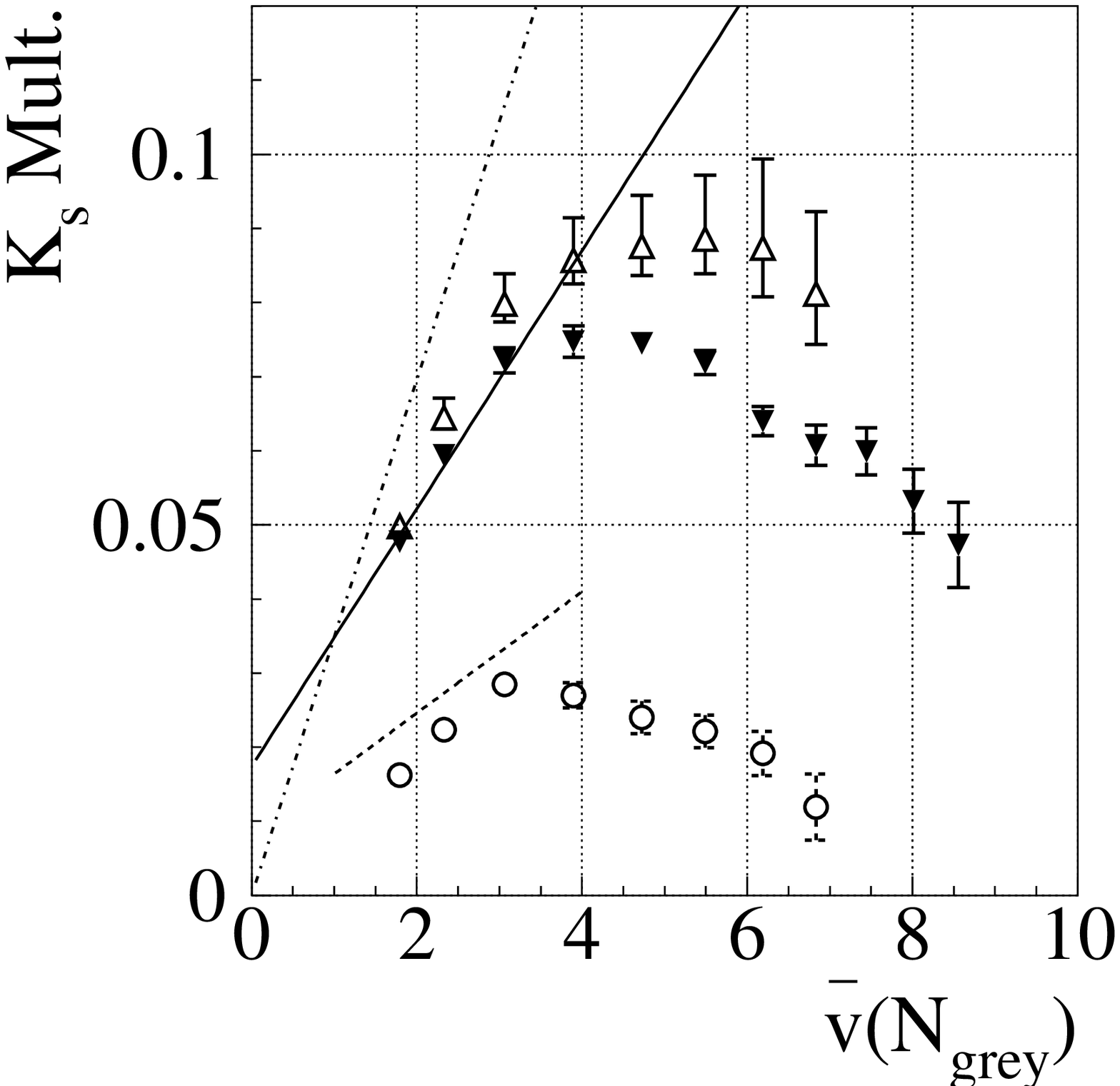,width=1.65in,bbllx=6pt,bblly=5pt,bburx=491pt,bbury=488pt} 
\end{tabular}
}
\caption{\Lam{} (left) and \Ks{} (right) multiplicities vs
\Avgnungrey. $\blacktriangledown$ - in E910 acceptance, $\vartriangle$
- total yield (syst. errors from extrapolation included),
$\circ$ - est. \Kzbar\ component of 
\Ks. Lines: Solid - \Npart\ scaling, dot-dashed - Binary
collision, dashed - empirical fit to \Lam\ (left panel, see text for
details), est. \Npart\ scaling for \Kzbar\ (right panel).
}
\label{fig:yield}
\end{figure}

We observe that with increasing \Ngrey, the \Lam{} and \Ks{} yields decrease
at high rapidity and increase at low rapidity. For \Lam's, the
decrease in yield 
at large rapidity is a direct consequence of the increased
``stopping'' of the projectile baryon resulting from the multiple
interactions in the target nucleus \cite{Chemakin:1999cp}. 
The strong increase in production of strange particles at
low rapidity is qualitatively consistent with previously observed
trends in secondary particle production in \pA{} collisions 
\cite{Abb92:Measurement,Eli80,Dem84:PA}. We show in Fig.~\ref{fig:yield}
the integrated yields that we have obtained by summing our measured
\dndy\ values as a function of \Nnu. The error on the in-acceptance
yield is dominated by and taken to be the same as our uncertainty in
the absolute normalization $\pm 10\%$ at $90\%{\rm CL}$. We show (solid
line) in Fig.~\ref{fig:yield} the expectations from Eq.~\ref{eq:wn} with
$N_{pp}^{\Lambda} = 0.054 \pm 0.002 (syst)$ and $N_{pp}^{K_s} = 0.035
\pm 0.002(syst)$ obtained by parameterizing the $\sqrt{s}$ dependence
of \Lam\ and \Ks\ multiplicities \cite{Gazdzicki:1996pk} and
interpolating to our energy. 
%%
%%($\sqrt{s} = 5.9$~GeV). 
%%
The \Lam{} yields initially increase faster with \Nnu\ than expected
from the \Npart\ scaling of \pp\ yields and then saturate and start to
decrease. The \Ks{} yields behave similarly with a slower initial increase.
The apparent decrease of the yields may result from the fact 
that we miss a larger fraction of the total \Lam{} and \Ks{} yield
with increasing \Ngrey{} or \Nnu{} due to our low-rapidity cut-off. We
have estimated the missing yield by 
fitting the \dndy{} distributions to gamma distributions  
%%%\cite{gammanote} 
as shown in Fig.~\ref{fig:yield} and extrapolating
these into the unmeasured region to 
produce the estimated total yields shown in Fig.~\ref{fig:yield}.
The uncertainty in the total yield is largest for the larger \Ngrey{} or
\Nnu{} bins where the \dndy{} distribution peaks near the edge of our
acceptance.  We show in Fig.~\ref{fig:yield} $90\% \rm{CL}$ systematic
errors on the yields with larger errors on the high side to account
for the possibility that an unknown mechanism may produce a larger
\Lam\ yield below $y=0$ than we estimate. The resulting total \Lam{}
yields shown in Fig.~\ref{fig:yield} saturate at large \Nnu{} and
remain flat. We have fit the extrapolated \Lam{} yields to an
empirical function, 
\begin{equation}
\label{eqn:yieldfit}
N_{\Lambda} = N^{\Lambda}_{pp} \, 
(1 - e^{-\kappa\,\nu^\alpha}) / (1 - e^{-\kappa}),
\end{equation}
where $N_{pp}$ is the \Lam{} multiplicity in \pp{} collisions. The
obtained function with $\kappa = 0.299\pm{}0.008$ and $\alpha =
1.29\pm{}0.03$ describes both the initial rapid rise of the yield and
the saturation at large \Nnu.
To evaluate the significance of this fast initial increase in the
\Lam{} yield we plot in Fig.~\ref{fig:yield} the yield that would result
from a ``binary-collision'' scaling of \pp{} data,
\begin{equation}
\label{eqn:bcscaling}
N^{\rm BC}(\nu) = N_{pp} \, \nu,
\end{equation}
which we view as the fastest plausible increase that could be
expected from the multiple scattering of the incoming proton. The
\Lam{} yields are consistent with this ``upper limit'' for 
$\nu \leq 3$ indicating a rate of increase in \Lam\ yield with \Nnu\
in this region that is approximately twice that given by Eq.~1.
We note that the systematic error on the \Nnu\ scale of $\pm 15\%$ 
\cite{Che99:Slow-Proton} is small compared to this difference in slope.

We observe that the \Ks\ yield increases more slowly with \Nnu\ than
the \Lam\ yield and appears to decrease slightly at
large \Nnu{} even after we have accounted for the missing yield. The
difference in behavior between the \Lam{} and \Ks{} yields may result
from the mixture of \Kz\ and \Kzbar\ in the \Ks\ and the fact that
these are produced through different processes. In \pp\ collisions at
comparable energies, $\frac{1}{2}$ of the \Ks\ are produced as \Kz's in
association with hyperons with the other half produced as \Kzbar's
associated with kaons \cite{Alp76:strange}. If we assume that this
proportion is not modified in \pA\ collisions and that the total
hyperon yields increase in proportion to the \Lam\ yield,
we can estimate the \Kzbar\ component of the measured \Ks\ yields shown
in Fig.~\ref{fig:yield}. The \Kzbar\ component appears to increase by
a factor of two at $\nu = 3$, roughly consistent with the \Npart\ scaling of
\pp\ data shown in Fig.~\ref{fig:yield}, before starting to
decrease slowly with \Nnu. While our data suggests an increase in
\kkbar\ production with \Nnu, because of uncertainties in the above
assumptions we cannot make a stronger statement. A forthcoming
analysis of \Km\ production will provide clearer insight into this problem. 

In conclusion, we have reported on the first detailed investigation of
the centrality dependence of strange particle production in \pA\
collisions using grey track multiplicity and the estimated number of
collisions of the projectile nucleon to characterize centrality. We
have measured \dndy\ distributions for \Lam\ and \Ks\ that both show a
strong backward shift with increasing \Ngrey\ and \Nnu. The estimated total
\Lam\ yields increase with \Nnu\ at a rate 
approximately twice that expected from the \Npart\ scaling of \pp\
data for $\Nnu \leq 3$ and saturate for $\nu >5$. As noted above, this
violation of \Npart\ scaling implies either that additional nucleons
not directly struck by the projectile contribute to \Lam\ production
or that the probability for one or more of the participants to
fragment into a \Lam\ increases with \Nnu. We observe a slower but
significant increase in \Ks\ multiplicity with \Nnu\ that apparently
results from different behavior of the \Kz\ and \Kzbar\ components of
the \Ks. The observed increase in \Ks\ yield is large enough to allow
for a statistically significant increase in \kkbar\ production with
\Nnu\ for $\nu \leq 3$ using a reasonable extrapolation of \pp\ data.

We conclude that at AGS energies, \pA\ data show a clear violation of
a simple \Npart\ scaling of \pp\ data. This result has clear
qualitative implications for use of such scaling for interpreting 
strangeness yields in A-A collisions. To quantitatively evaluate the
potential implications of our results, we assume that the target
contribution to the \pAu\ \Lam\ yield grows as $\nu N_{pp}/2$ and
attribute the remainder to the fragmentation of the projectile and/or
energy deposition of the projectile in the nucleus. Then, our data
show that the ``projectile'' contribution increases proportional to
$\nu$ for $\nu \leq 3$ with a slope that is the same as for the target 
nucleons. In \mbox{A-A} collisions where {\it both} the projectile and
target nucleons multiply scatter, this picture implies that the
hyperon and associated kaon yields per participant would increase
rapidly with the average number of scatters of the participants,
$\langle \nu \rangle$, for $\langle \nu \rangle \leq 3$ giving a
maximum possible increase in yield per participant of a factor $\sim
3$. This is precisely the behavior seen in the \Kp\ production
in \SiAu\ and \AuAu\ collisions at the AGS \cite{Ahle:1999va} and
\Lam\ production in \PbPb\ collisions at the
SPS\cite{Andersen:1999ym}. In \pA\ 
collisions, the enhancement is more modest -- a 50\% increase over
\Npart\ scaling at $\nu = 3$ -- simply because the target nucleons
scatter only once. As we have shown, however, this modest enhancement
may have profound consequences for interpretation of the strangeness
enhancement in nuclear collisions. We note that the above picture
picture is consistent with the additive quark model
\cite{Bialas:1977en} used by Kadija~\etal \cite{Kadija:1995fa} to
explain the observed strangeness enhancement in light-ion collisions
at the SPS. In particular, the increase in the
projectile-like component up to $\nu = 3$ is exactly what is expected
from the additive quark model. Since the saturation of the \Lam\ yield
for $\Nnu >5$ is very likely due to the stopping of the incident
baryon we predict that at higher energies the \Lam\ yield will
continue to increase for $\Nnu > 5$.  

We wish to thank Dr. R.~Hackenburg and the MPS staff, J.~Scaduto and
Dr. G.~Bunce. This work has been supported by the U.S. Department of Energy 
under contracts with BNL (DE-AC02-98CH10886), Columbia 
(DE-FG02-86ER40281), ISU (DOE-FG02-92ER4069), KSU (DE-FG02-89ER40531), 
LBNL (DE-AC03-76F00098), LLNL (W-7405-ENG-48), ORNL (DE-AC05-96OR22464)
and UT (DE-FG02-96ER40982) and the National
Science Foundation under contract with FSU (PHY-9523974).

\bibliography{journals,hi,notes}

\begin{thebibliography}{10}
\bibitem{npart}
The projectile counts as one participant no matter how many times it scatters.
\bibitem{accept_note}
We have neglected small ($< 5\%$) multi-track losses.
\end{thebibliography}
\bibliographystyle{prsty}

%% $Source: /users/usr0/e910/e910root/papers/cvssrc/lambdaKsPRL/tex/body.tex,v $

% \begin{table}[t]
% \begin{tabular}{ l c c c c }
%        & $N_0$ & $\kappa$ & $\alpha$ & $\chi^2/dof$ \\\hline
% \Lam{} & 0.201 & 0.245    & 1.61     & 10.6/6       \\
% \Ks{}  & 0.089 & 0.285    & 1.81     &  1.5/4       \\
% \end{tabular}
% \caption{Results of fitting \Lam{} and \Ks{} estimated total yield 
% with Eq.~\ref{eqn:yieldfit}.}
% \label{tab:yieldfit}
% \end{table}

\end{document}